\documentclass[preprint,showpacs,preprintnumbers,amsmath,amssymb]{revtex4}

\usepackage{graphicx}
\usepackage{dcolumn}
\usepackage{bm}

\begin{document}


\title{ Systematic effects of carbon doping on the
superconducting properties of Mg(B$_{1-x}$C$_x$)$_2$}

\author{R. H. T. Wilke$^{\dagger\ddagger}$}
\author{S. L. Bud'ko$^{\dagger}$}%
\author{P. C. Canfield$^{\dagger\ddagger}$}%
\author{D. K. Finnemore$^{\dagger\ddagger}$}%
\email{finnemor@ameslab.gov} \affiliation{
$^{\dagger}$Ames Laboratory and $^{\ddagger}$Department of Physics and Astronomy\\
Iowa State University, Ames, IA 50011}
\author{Raymond J. Suplinskas}
\affiliation{Specialty Materials, Inc., 1449 Middlesex Street,
Lowell, MA 01851}
\author{S. T. Hannahs}
\affiliation{National High Magnetic Field Laboratory, Florida State University\\
1800 E. Paul Dirac Drive, Tallahassee, FL 32310}

\date{\today}

\begin{abstract}

The upper critical field, $H_{c2}$, of Mg(B$_{1-x}$C$_x$)$_2$ has
been measured in order to probe the maximum magnetic field range
for superconductivity that can be attained by C doping. Carbon
doped boron filaments are prepared by CVD techniques, and then
these fibers are then exposed to Mg vapor to form the
superconducting compound. The transition temperatures are
depressed about $1~K/\%$ C and $H_{c2}(T=0)$ rises at about
$5~T/\%$ C.  This means that $3.5\%$ C will depress $T_c$ from
$39.2~K$ to $36.2~K$ and raise $H_{c2}(T=0)$ from $16.0~T$ to
$32.5~T$.  Higher fields are probably attainable in the region of
$5\%$ C to $7\%$ C.  These rises in $H_{c2}$ are accompanied by a
rise in resistivity at $40~K$ from about $0.5~\mu \Omega~cm$ to
about $10~\mu \Omega~cm$.  Given that the samples are
polycrystalline wire segments, the experimentally determined
$H_{c2}(T)$ curves represent the upper $H_{c2}(T)$ manifold
associated with $H\perp c$.

\end{abstract}



\maketitle

The discovery of superconductivity with $T_c~\sim ~40~K$ \cite {1}
focused a lot of attention on the previously ignored compound,
MgB$_2$.  Over the past few years, a great deal has been learned
about high purity samples of this material, ranging from the
anisotropic nature of the upper critical field, $H_{c2}$, to the
two gap nature of the superconducting state.\cite {2}  On the
other hand, despite much effort, systematic studies of the
properties of doped MgB$_2$ have been made difficult by the fact
that, under more common reaction routes that often involve
diffusing Mg vapor into the B matrix at or near atmospheric
pressure, it is difficult to homogenously substitute atoms for
either the Mg or B sites.\cite {3}  Despite this uncertainty about
the distribution of impurity atoms, it has been possible to raise
$H_{c2}$ by a factor of 2 or more with the upper critical field at
$T=0$, $H_{c2}(0)$, rising from $16~T$ for pure samples to $\sim
30~T$ or even $\sim 48~T$ in "dirty" thin films.\cite {4} These
reports clearly indicate that there is very likely a way of
judiciously doping MgB$_2$ to increase $H_{c2}$ dramatically.

Recently it has been demonstrated that carbon can be uniformly
substituted for boron if the carbon and boron are mixed at an
atomic length scale by using B$_4$C as the starting material.\cite
{5} The resulting Mg(B$_{1-x}$C$_x$)$_2$ material \cite{5,6} had
$x=0.10\pm0.02$ and a sharp superconducting transition at $22~K$.
The remaining C goes into an MgB$_2$C$_2$ phase. Even with this
strongly suppressed transition temperature, the $x\approx 0.10$
Mg(B$_{1-x}$C$_x$)$_2$ sample manifests clear evidence of two gap
superconductivity \cite {5,7} and had $H_{c2}(0)\sim 25~T$, \cite
{8} a value that exceeds the $16~T$ $H_{c2}(0)$ of pure MgB$_2$.
These data strongly supported the idea that carbon may well be a
key dopant for tuning $H_{c2}$ in bulk and perhaps also in thin
films.

It is clearly desirable to systematically study the effects of low
carbon concentration on the superconducting properties of MgB$_2$,
but the C-B binary phase diagram indicates that B$_4$C is the most
B-rich compound. Even though there is a substantial width of
formation for B$_4$C, the smallest amount of C that can be present
in equilibrium in a boron/carbon binary compound is about $10\%$
C.\cite {9}  In order to study the systematic effects for $x\le
0.1$ in Mg(B$_{1-x}$C$_x$)$_2$, some non-equilibrium method of
intimately mixing B and C must be used.  Fortunately wire segments
of MgB$_2$ can be synthesized from boron filaments made by a
chemical vapor deposition (CVD) process similar to the methods
used to make commercial boron filaments in kilometer lengths.\cite
{10} Carbon can be co-deposited with the boron in a controlled
manner by introducing CH$_4$ in the BCl$_3$ and H$_2$ gas streams
used in the CVD fiber processing.

In this work, we present a systematic study of the changes in
$H_{c2}$ that occur in Mg(B$_{1-x}$C$_x$)$_2$.  For the addition
of C up to $x\sim 0.035$ by this CVD method the transition
temperature, $T_c$, is suppressed only slightly, from $39.2~K$ to
$36.2~K$, whereas the $H_{c2}(0)$ is increased from $16~T$ to
$32~T$.  With these data, as well as our earlier data on $x\sim
0.10$ \cite {5,6,7} we can tentatively explain the $H_{c2}(0)$
values seen in some thin film samples, even an $H_{c2}(0)$ of
$\sim 48~T$.  These data indicate that not only does
Mg(B$_{1-x}$C$_x$)$_2$ have a $T_c$ that vastly exceeds other
intermetallic superconductors such as Nb$_3$Sn, \cite {11} but it
also can have $H_{c2}(0)$ values that exceed Nb$_3$Sn's upper
critical field performance.

To prepare a carbon doped B fiber, a W wire about $15~\mu m$ in
diameter is passed through a Hg seal at the rate of a few $cm/s$
into a long glass tube.\cite {10} A flowing gas stream of BCl$_3$,
H$_2$, and CH$_4$ move through the full length of the chamber. The
fiber is heated electrically to temperatures in the
$1100-1300^\circ C$ range, and the boron and carbon are deposited
together to form a fiber of about $75~\mu m$ diameter and a few
hundred meters length.  Nominal ratios of C to B in the gas stream
were selected to be zero, $0.5\% $, $1\% $, and $2\% $ and several
hundred meters of fiber were made. Short lengths of fiber are
placed in a Ta tube with excess Mg \cite {12} with about a Mg/B
ratio of 1. The sealed Ta tube is again sealed in quartz and
placed in a box furnace and heated to about $1200^{\circ }C$.
Preparation using a steady ramp from $650^{\circ }$C to
$1200^{\circ }$C gave the best flux pinning and these are the
samples reported here. One of the unexpected aspects of sample
preparation was that $C$ doping, even at the lowest level studied,
substantially slows the formation of the Mg(B$_{1-x}$C$_x$)$_2$
phase. Whereas pure B converts completely to MgB$_2$ in the
presence of Mg vapor at $950^{\circ }C$ for $2~h$, the boron doped
with $0.5\%$ C fiber requires about $1200^{\circ }C$ for $48~h$ to
transform to Mg(B$_{1-x}$C$_x$)$_2$. Short wire segments of
Mg(B$_{1-x}$C$_x$)$_2$ are removed from the Ta container and lead
wires are attached with Ag epoxy.

X-ray patterns for each of the reacted Mg(B$_{1-x}$C$_x$)$_2$
samples show the MgB$_2$ phase, Mg lines, and the fiducial Si
lines that were used to calibrate the $2\Theta $ angle
measurements. A search was made for the obvious impurity lines,
but no evidence of MgO or B$_4$C are present. Fig. 1a shows an
expanded view of the $(002)$ peak and the $(110)$ peak of the
Mg(B$_{1-x}$C$_x$)$_2$ phase. Within the accuracy of these
measurements, there is no shift in the $(002)$ peak indicating
that there is no measurable change in the c-axis lattice
parameter.  The $(110)$ peak shifts systematically from $59.85$
degrees for the pure sample to $60.10$ degrees for the nominal
$2\% $ sample. Using these measured changes in the a-lattice
parameter we can determine the amount of carbon in our samples via
comparison to $\Delta a(x)$ from the Lee et al. Auger electron
spectroscopy (AES) analysis \cite {13} as well as the Avdeev et
al. neutron diffraction data.\cite {6} This comparison is shown
graphically in Fig. 1b and indicates that our
Mg(B$_{1-x}$C$_x$)$_2$ samples have $x=0.0095$, $x=0.017$, and
$x=0.035$.

Carbon doping raises the the resistivity at $40~K$, $\rho (40K)$
from the pure MgB$_2$ value of $0.5\pm 0.2\mu \Omega~cm$ to the
$x=0.035$ value of $10\pm 4~\mu \Omega~cm$.  This uncertainty may
arise from the rather short distance between voltage contacts on
the sample, less than $1~mm$,  and the different time temperature
profiles in making the Mg(B$_{1-x}$C$_x$)$_2$ samples. In order to
address this problem, a series of three samples of the $x=0.035$
material prepared with a ramp from $650^{\circ }$C to $1200^{\circ
}$C with somewhat longer distance between voltage contacts were
measured and the $\rho (40K)$ values in zero field were found to
be $9.9, 10.1$, and $10.6~\mu \Omega cm$.

The $T_c$ values, shown in Fig. 2 for each x-value, were measured
both from magnetization and resistivity vs. temperature sweeps as
shown by the two insets. With a measuring field of  $50~Oe$,
magnetization always shows about $60-80~G$ Meissner screening at
$5~K$. The data are plotted on the inset as values normalized by
the $5~K$ magnetization. The magnetization $T_c$ is defined by the
onset of flux exclusion. The resistive $T_c$ is defined by an
onset criterion (extrapolation of maximum slope up to normal state
resistivity) so as to be consistent with the criterion used below
for higher field data. For these measurements a current density of
$\sim 6~A/cm^2$ was used. Values of $T_c$ derived from
magnetization (solid circles) and resistivity (solid squares) are
plotted on Fig. 2. The initial slope is about $1~K/\%$ C in the
$x=0$ to $x=0.035$ range.

The $H_{c2}$ values have been determined from both $\rho $ vs $T$
data up to $14~T$ taken on a Quantum Designs PPMS  and from $\rho
$ vs. $H$ sweeps taken at the National High Magnetic Field
Laboratory.  Results for the $x=0.035$ sample are shown in Fig. 3a
and 3b respectively. The resistivity vs field sweeps show a
classic zero resistance range up to a critical depinning current
followed by a linear $\rho $ vs. $H$ characteristic of flux flow
resistivity. \cite {14}  It should be noted that given the
polycrystalline nature of our samples, our measurements of
$H_{c2}(T)$ are determinations of the uppermost $H_{c2}(T)$ curve
which is $H_{c2}^{\perp c}(T)$ for the pure compound. \cite
{2,15,16}

The central result of this work is shown in the $H_{c2}$ vs. $T$
plots of Fig. 4.  The data for two different $x=0.035$ samples
shown by the solid and open squares are quite linear and rise well
above a Werthamer, Helfand, and Hohenberg prediction \cite {18} of
$H_{c2}(0)=24.5~T$ determined by fitting the slope in the $20$ to
$30~K$ range.  To show the behavior near $T_c$, an inset on Fig. 4
shows the behavior of the $x=0.035$ sample with both $R$ vs. $H$
data (solid triangles) taken at the National High Magnetic Field
Laboratory and $R$ vs. $T$ data (solid squares) taken on the PPMS
up to $14~T$. Magnetization data ($M$ vs. $H$) is reversible near
$T_c$ and show a very clean change in slope when the sample begins
to expel flux.  These results were taken down to $30~K$ and are
shown by the open squares.  These three methods to determine
$H_{c2}$ are very consistent especially when it is noted that each
was made with different samples.  Results for the $x=0.017$ sample
is similar to the $x=0.035$ sample except that the $T_c$ is higher
at $37.9~K$ and $H_{c2}(0)=25~T$.  For $x=0.0095$, the values are
$T_c=38.6~K$ and $H_{c2}(0)=20~T$.  The $H_{c2}$ vs. $T$ curves,
all of these samples show positive curvature near $T_c$ and
negative curvature near $T=0$ with a rather linear behavior over
much of the temperature range.

For standard type II superconductors decreases in the electronic
mean free path directly manifest themselves as increases in
$H_{c2}(T)$.  In the case of MgB$_2$ though, even a change in the
residual resistitivity of an order of magnitude may not be enough
to place this sample into the dirty limit ($l \ll \xi_0$).  These
statements, though, are based on single band, single scattering
time arguments which are very likely to be incorrect or incomplete
for MgB$_2$, given its two bands and gaps and at least three
scattering times (two intraband scattering times and one interband
scattering time).  Gurevich \cite {19} has made predictions about
the form and size of $H_{c2}(T)$ for MgB$_2$ in the dirty limit,
but this model requires assumptions about all three scattering
times. The data we present in Fig. 4 along with our measured
changes in resistivity provide points of reference for this
theory, i.e. for an order of magnitude increase in $\rho_0$ we
double $H_{c2}(0)$, but still have upward curvature near $T_c$
(Fig. 4 inset) and also have a clear roll-over at low
temperatures.

To summarize our results and place them in context with other data
on Mg(B$_{1-x}$C$_x$)$_2$ we have plotted both $T_c$ and
$H_{c2}(0)$ as a function of carbon content for our own data as
well as selected experiments on single crystals on Fig. 5.   The
carbon content for the data by Avdeev et al. \cite {6} was
determined by neutron diffraction whereas Lee et al. \cite {13}
used AES and Kazakov et al. \cite {18}, as well as the present
work used the shift in the $a$-lattice parameter.  The first point
that is worth noting is that the $T_c(x)$ manifold, which contains
data from 4 different groups on samples synthesized by a variety
of ways (low and high pressure synthesis) and in a variety of
forms (small single crystals, bulk wire segments and sintered
pellets), is quite reproducible and robust. This is very
encouraging and indicates that $T_c$ may be used as a rough
caliper of how much carbon is in a given sample. Focusing on the
$H_{c2}(0)$ of this work at low C concentration (filled circles),
the rapid rise of $H_{c2}(0)$ at $\sim 5~T/\%$ C would seem to
indicate that values in the $45~T$ range may well be possible. On
the other hand the measurements of $H_{c2}(0)=25~T$ for a carbon
content of about $x\sim 0.10$, \cite {8} indicate that the initial
rapid rise of $H_{c2}(0)$ seen for lower C contents will
eventually bend over and reach a maximum at intermediate carbon
concentrations. The region from $x=0.04$ to $x=0.07$ is clearly of
interest. Whereas these data show a systematic increase in
$H_{c2}(0)$ for $x\le 0.035$ carbon doping levels, they also
indicate that higher $H_{c2}(0)$ values clearly should be
anticipated for slightly higher $x$-values. Indeed, one of the
largest $H_{c2}(0)$ values reported to date ($H_{c2}(0)\sim 47~T$
for a thin film with a $T_c\sim 31~K$ \cite {4}) is not
inconsistent with a linear extrapolation of the $H_{c2}(0)$ line
for $x\le 0.035$ from this work to larger values of $x$. This is
consistent with the $T_c$ value of the film implying a carbon
content in the vicinity of $x \approx 0.07$. This strongly
suggests that carbon may have been the dominant impurity changing
$H_{c2}(0)$ in this film. All of these data taken together
strongly suggest that with judicious carbon doping,
Mg(B$_{1-x}$C$_x$)$_2$ can be tuned to have remarkably large upper
critical field values, making this already fascinating material of
even greater interest.

\begin{acknowledgments}
Ames Laboratory is operated for the US Department of Energy by
Iowa State University under Contract No. W-7405-Eng-82. This work
was supported by the Director for Energy Research, Office of Basic
Energy Sciences. A portion of this work was performed at the
National High Magnetic Field Laboratory, which is supported by NSF
Cooperative Agreement No. DMR-0084173 and by the State of Florida.
\end{acknowledgments}

\eject

\clearpage

\begin{figure}
\begin{center}
\includegraphics[angle=0,width=95mm]{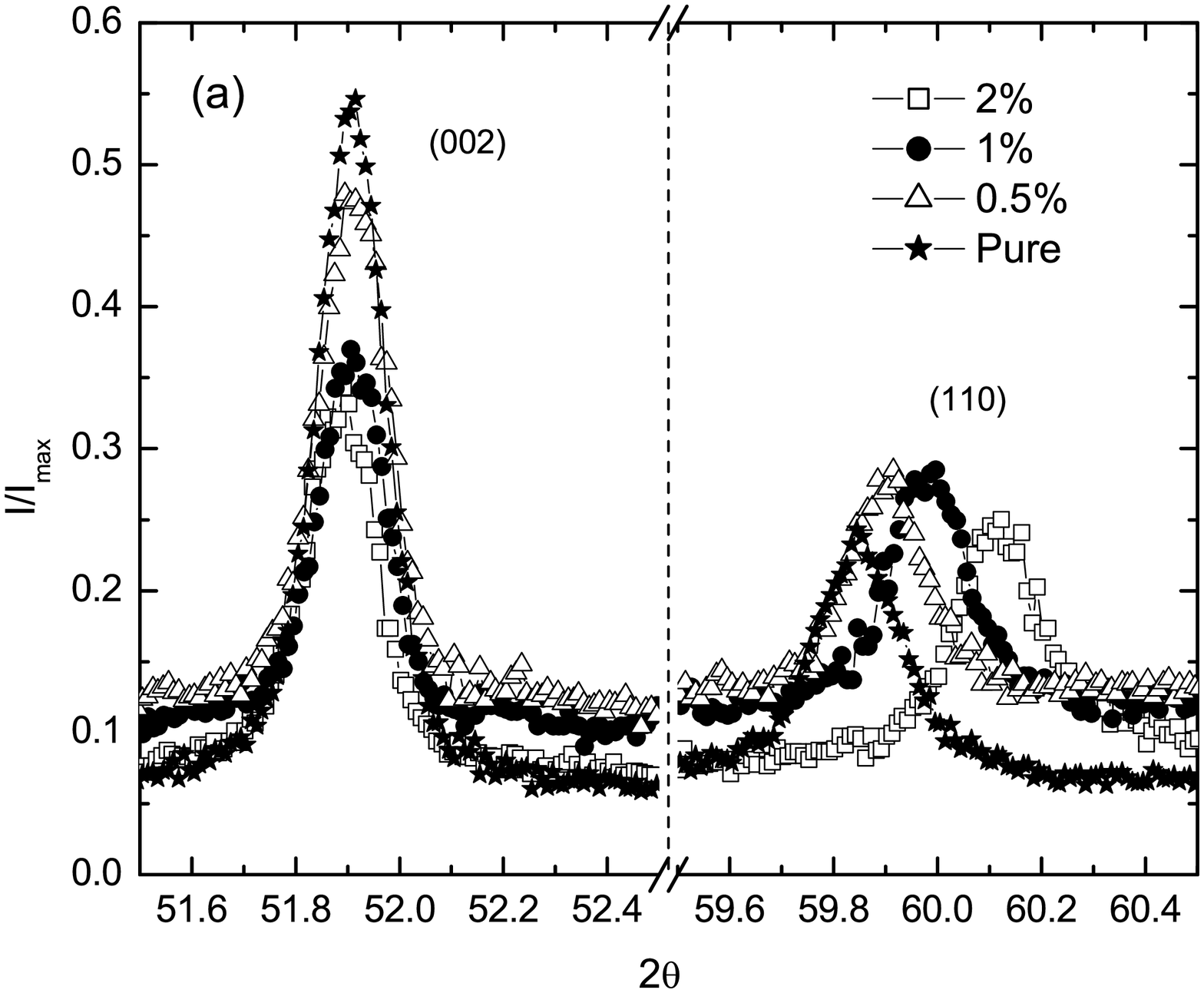}
\includegraphics[angle=0,width=95mm]{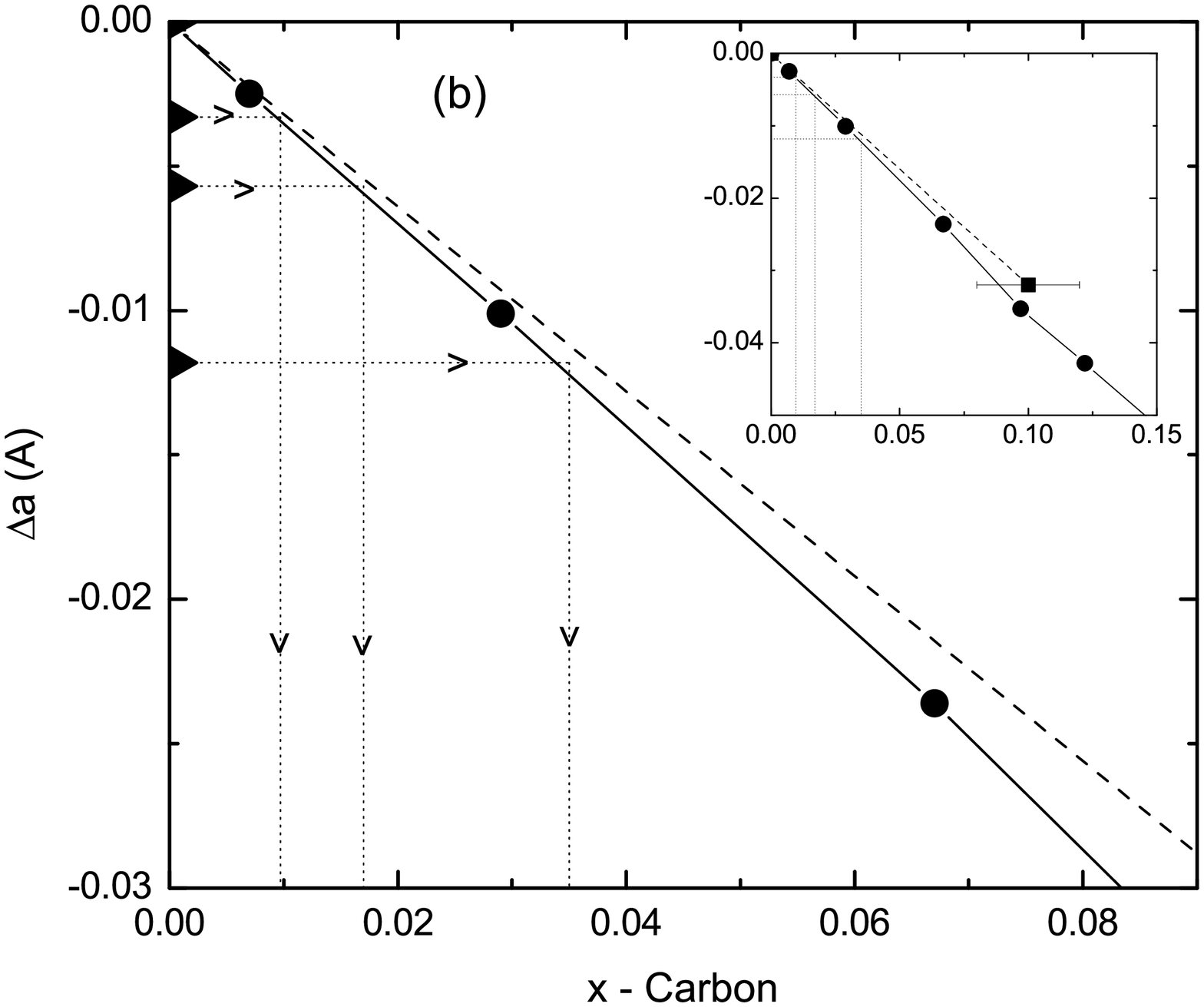}
\end{center}
\caption{(a) selected regions of powder X-ray diffraction data
from Mg(B$_{1-x}$C$_x$)$_2$ wire segments synthesized from boron
filaments with $0\%$, $0.5\%$, $1\%$, and $2\%$ nominal carbon
substitution.  Note that whereas there is no detectable shift of
$(002)$ peak, $(110)$ shifts systematically with carbon
substitution. (b) Shift in $a$-lattice parameter as a function of
carbon content from neutron diffraction data [6] show as dashed
line and AES data [13] shown as solid line with our shift in
$a$-lattice parameter shown as large triangles on y-axis.
Projection of our $\Delta a$ data onto the $\Delta a(x)$ lines
(shown as dotted lines) indicate that the our three carbon
substituted Mg(B$_{1-x}$C$_x$)$_2$ samples have $x \approx$
0.0095, 0.017, and 0.035. Inset shows data for an enlarged carbon
range, $x < 0.15$.}\label{F1}
\end{figure}

\clearpage

\begin{figure}
\begin{center}
\includegraphics[angle=0,width=120mm]{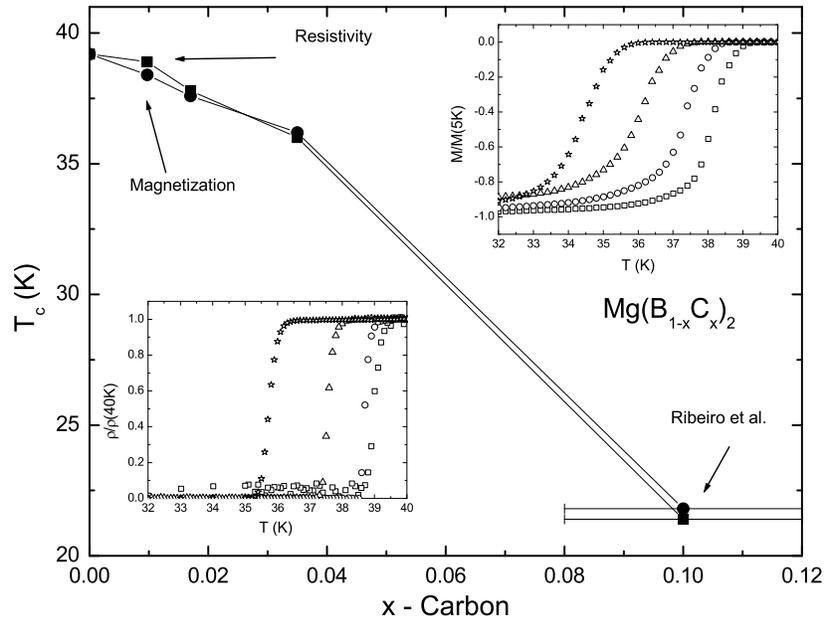}
\end{center}
\caption{Superconducting transition temperature as a function of
carbon content.  Data inferred from temperature dependent
resistivity (shown in lower inset) shown as squares and data
inferred from temperature dependent, low field magnetization
(shown in upper inset) shown as circles. Data from Ribeiro et al.
[5] for $x \approx .10$ are also shown. For insets squares
indicate $x = 0$, circles - $x = 0.0095$, triangles - $x = 0.017$,
stars - $x = 0.035$.}\label{F2}
\end{figure}

\clearpage

\begin{figure}
\begin{center}
\includegraphics[angle=0,width=120mm]{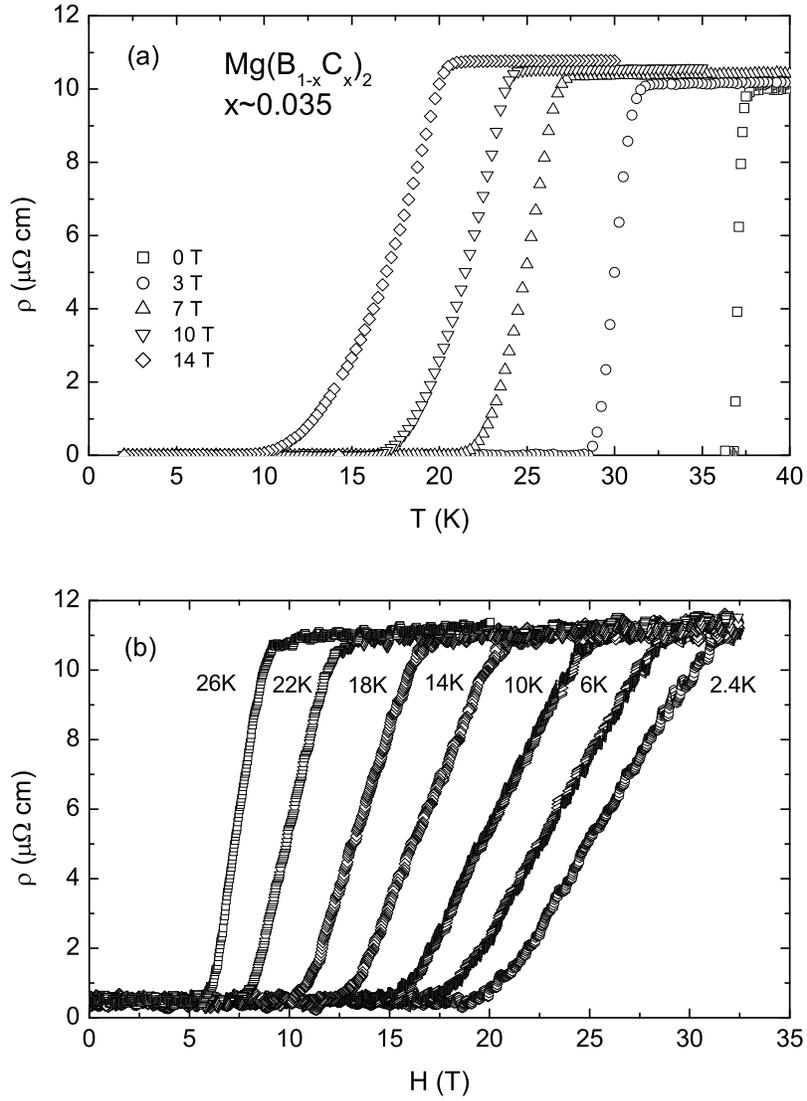}
\end{center}
\caption{ Temperature (a) and magnetic field (b) dependent
electrical resistivity of Mg(B$_{1-x}$C$_x$)$_2$ with $x = 0.035$.
The data presented in (a) are for $H \leq 14 T$ and taken in a
Quantum Design PPMS system.  The data in (b) are for $H \leq 33 T$
and were taken at the NHMFL, Tallahassee.}\label{F3}
\end{figure}

\clearpage

\begin{figure}
\begin{center}
\includegraphics[angle=0,width=120mm]{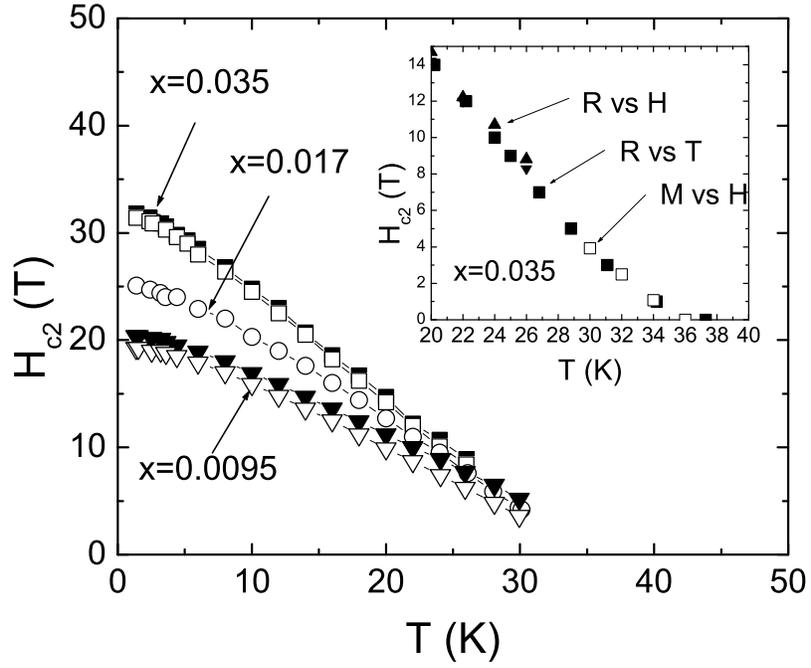}
\end{center}
\caption{Superconducting upper critical field, $H_{c2}$, as a
function of temperature for Mg(B$_{1-x}$C$_x$)$_2$, $x \leq 0.035$
samples. Inset: $H_{c2}(T)$ closer to $T_c$ determined from
temperature dependent resistivity (filled square), field dependent
resistivity (triangle) and field dependent magnetization (open
square).}\label{F4}
\end{figure}

\clearpage

\begin{figure}
\begin{center}
\includegraphics[angle=0,width=120mm]{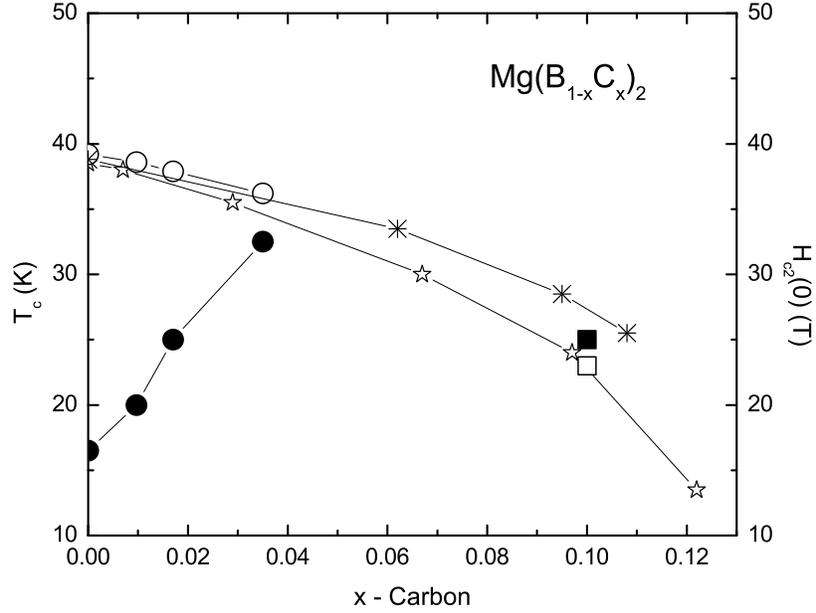}
\end{center}
\caption{Superconducting upper critical field, $H_{c2}$, and
superconducting transition temperature, $T_c$, as a function of
$x$ for Mg(B$_{1-x}$C$_x$)$_2$ samples.  $T_c(x)$ data are (open
circles) this study, (open square) $x \approx 0.10$ from Ribeiro
et al.[5], (stars) and (asterisks) for single crystal samples from
Lee et al. [13] and Kazakov et al. [18] respectively. $H_{c2}(0)$
data are (filled circles) this work and (filled square) $x \approx
0.10$ from Holanova et al.[8]. It should be noted that (i) all of
the $T_c(x)$ data agree quite well and (ii) that there clearly
will be a maximum $H_{c2}(0)$ for $0.035 < x < 0.10$ that can be
expected to be between $\sim 35$ and $\sim 55 T$.}\label{F5}
\end{figure}

\end{document}